\documentclass[aps,prl,preprint,groupedaddress,showpacs]{revtex4-1}
\usepackage{graphicx}
\begin{document}

% Use the \preprint command to place your local institutional report
% number in the upper righthand corner of the title page in preprint mode.
% Multiple \preprint commands are allowed.
% Use the 'preprintnumbers' class option to override journal defaults
% to display numbers if necessary
%\preprint{}

%Title of paper
\title{Derivation of the relativistic momentum and relativistic equation of motion from Newton's second law and Minkowskian space-time geometry}

% repeat the \author .. \affiliation  etc. as needed
% \email, \thanks, \homepage, \altaffiliation all apply to the current
% author. Explanatory text should go in the []'s, actual e-mail
% address or url should go in the {}'s for \email and \homepage.
% Please use the appropriate macro foreach each type of information

% \affiliation command applies to all authors since the last
% \affiliation command. The \affiliation command should follow the
% other information
% \affiliation can be followed by \email, \homepage, \thanks as well.
\author{Krzysztof R\c ebilas}
\email[]{krebilas@ar.krakow.pl}
%\homepage[]{Your web page}
%\thanks{}
%\altaffiliation{}
\affiliation{Zak\l{}ad Fizyki. Akademia Rolnicza im. Hugona Ko\l{}\l{}\c ataja. 
Al. Mickiewicza~21, \mbox{31-120} Krak\'ow. Poland.
 Telephone: \mbox{48-12-6624390}. Fax: \mbox{48-12-6336245}.
}

%Collaboration name if desired (requires use of superscriptaddress
%option in \documentclass). \noaffiliation is required (may also be
%used with the \author command).
%\collaboration can be followed by \email, \homepage, \thanks as well.
%\collaboration{}
%\noaffiliation

\date{\today}

\begin{abstract}
Starting from the classical 
Newton's second law which, according to our assumption,  is valid in any instantaneous inertial rest frame of   body that moves in Minkowskian
 space-time we get the relativistic equation of motion $\vec{F}=d\vec{p}/dt$, where $\vec{p}$ is the relativistic momentum.  The relativistic momentum  is then  derived without referring to any additional assumptions concerning elastic collisions of bodies.  Lorentz-invariance of the relativistic law is proved without  tensor formalism. Some new method of force transformation is also presented.

\end{abstract} 

% insert suggested PACS numbers in braces on next line
% insert suggested keywords - APS authors don't need to do this
\maketitle

%\maketitle must follow title, authors, abstract, \pacs, and \keywords
\section{Introduction}

Only two assumptions: Newton's law and Minkowskian space-time geometry suffice, according to our method, to derive 
the relativistic momentum and  equation of motion (with its invariance).  
We believe that the presented in this paper deductive approach  to relativistic dynamics has some advantages over the standard one based on 
additionally introduced ad hoc postulates. It  may be interesting for physicists  aiming to understand the logical foundations of the form of the relativistic equation of motion  and its direct connection with the space-time geometry and the definition of force (Newton's law).

The main purpose of this  work is to reveal a possibility of deriving the relativistic momentum and the relativistic law of motion purely by means of considering the dynamics of an accelerating body. 
Assuming that the classical Newton's second law
 $\vec{F}_R=md\vec{u}/d \tau$ is valid in any instantaneous inertial rest frame $R$ of moving body we  get for a stationary system of coordinates $S$ the
 relativistic equation $\vec{F}=d\vec{p}/dt$, with the relativistic momentum $\vec{p}=m\vec{v}/(1-v^2/c^2)^{1/2}$. Also the general relation
 between the force $\vec{F}_R$ and the force $\vec{F}$ in the frame $S$ is derived. Finally, a new approach to the transformation of force between two arbitrary frames is presented.  
 The only relativistic assumption we need is that the space-time geometry is Minkowskian, so that  the Lorentz
 transformation of coordinates is applied in our reasoning.

There are several advantages of the method presented in this paper. The first is  {logical} one:
 we do not need to introduce any unnecessary postulates that are used in the standard approaches. The traditional way [1,2] to derive the relativistic momentum requires the assumption that the relativistic momentum is a quantity conserved in elastic collisions independently of  the inertial reference frame in which it is measured. This is quite non-trivial \emph{extra} assumption which in fact should be verified by appropriate experiments. 

What is more, if the relativistic momentum is introduced in the traditional way, a {second} \emph{separate} postulate is needed to establish how a force influences the change of  relativistic momentum. The proposed form of the relativistic equation of motion is $\vec{F}=d\vec{p}/dt$ because it fulfils two important requirements: a) it reproduces the Newtonian dynamics at the limit of small velocities, b) this law is Lorentz-covariant for the electrodynamic Lorentz force. But it is merely a postulate to undergo further experimental investigation. It is evident that the requirement a) may be satisfied by many other equations of motion that at the classical limit reduce to Newton's second law. In turn, the requirement b) refers solely to one kind of forces, i.e.  the electromagnetic ones. 
Instead  of introducing an unfounded assumption  we propose a more deductive method to arrive at the relativistic equation of motion. Starting 
 from the well-known and experimentally confirmed Newton's law we
show that \emph{for any kind of force} the space-time geometry enforces a moving
 body to undergo just the \emph{relativistic} law.

Furthermore, our attitude is quite \emph{general}. We do not have to consider any sophisticated and uncommon thought-experiments (i.e. specially arranged collisions with  very small deflection of particles). Let us note also that, as far as we know, in all cases the relativistic momentum is derived, authors examine only a collision of two \emph{identical} bodies. The generality of conclusions obtained in such a  way cannot be satisfactory. 

Moreover, according to our approach
\emph{invariance} of the relativistic equation of motion  for {any} kind of accelerating force appears to be obvious and indispensable and no tensor formalism is necessary to prove it. Let us notice also that invariance of the relativistic law of motion is shown in literature only  for the electromagnetic forces.

\section{Derivation of the relativistic momentum and equation of motion}

Consider  an accelerating body moving along a completely arbitrary trajectory.
Let in an instantaneous inertial rest frame $R$ of the body (i.e. momentarily co-moving with the body) a force $\vec{F}_R$ causes within a time interval $d\tau$ a change of its velocity $d\vec{u}$. The force $\vec{F}_R$ is a function depending on some features of the force-source and is determined by a force law established in the rest frame $R$.

Let from the point of view of a stationary inertial frame $S$ the body possesses a velocity $\vec{v}$. 
Due to the action of force the  velocity $\vec{v}$ changes in the frame $S$ in the direction parallel and perpendicular to $\vec{v}$ according to the following equations, respectively [3]: 
\begin{equation}
(dv)_{\parallel}=\frac{(du)_{\parallel}+v}{1+\vec{v} \cdot d\vec{u}/c^2}-v\simeq (du)_\parallel
(1-v^2/c^2),
\label{dvr}
\end{equation}

\begin{equation}
(d\vec{v})_\perp=\frac{(d\vec{u})_\perp\sqrt{1-v^2/c^2}}{1+\vec{v} \cdot d\vec{u}/c^2}\simeq (d\vec{u})_\perp\sqrt{1-v^2/c^2}.
\label{dvp}
\end{equation}

Because in the instantaneous rest frame $R$ the velocity of body $\vec{u}$ is zero, the relativistic equation of motion in the frame $R$ has the form of the classical Newton's law:
\begin{equation}
\vec{F}_R=md\vec{u}/d \tau
\label{weknewton} 
\end{equation}
Remembering that $d\tau=(1-v^2/c^2)^{1/2}dt$ and using Eqs (\ref{dvr}) and (\ref{dvp}) we can rewrite this equation of motion valid in the rest frame $R$
 separately for the direction parallel and perpendicular to the velocity $\vec{v}$:
\begin{equation}
(F_{R}){_\parallel}=m\frac{(du)_\parallel}{d\tau}=\frac{m}{(1-v^2/c^2)^{3/2}}\frac{(dv)_\parallel}{dt},
\label{row}
\end{equation}

\begin{equation}
(\vec{F}_{R}){_\perp}=m\frac{(d\vec{u})_\perp}{d\tau}=\frac{m}{1-v^2/c^2}\frac{(d\vec{v})_\perp}{dt}.
\label{prost}
\end{equation}

Because $(dv)_\parallel\equiv dv$, the equation (\ref{row}) is formally identical to:
\begin{equation}
(\vec{F}_{R}){_\parallel}=\frac{d}{dt}\left(\frac{mv}{\sqrt{1-v^2/c^2}}\right)\vec{n}=\frac{dp}{dt}\vec{n},
\label{Fr1}
\end{equation}
where $p$ is the value of the relativistic momentum of the body in the frame $S$. We have introduced  a vector $\vec{n}=\vec{v}/v$ to have this equation explicitly in  vector notation.

In turn, Eq. (\ref{prost}) can be written as:
\begin{equation}
(\vec{F}_{R}){_\perp}=\frac{p}{\sqrt{1-v^2/c^2}}\frac{d\vec{n}}{dt}
\label{Fp1}
\end{equation}
because:
\begin{eqnarray}
\frac{d\vec{n}}{dt}=\left(v\frac{d\vec{v}}{dt}-\frac{dv}{dt}\vec{v}\right)\frac{1}{v^2}=\left(v\frac{(d\vec{v})_\perp}{dt}+ v\frac{(d\vec{v})_\parallel}{dt}-\frac{dv}{dt}\vec{v}\right)\frac{1}{v^2}\nonumber\\
=\left[v\frac{(d\vec{v})_\perp}{dt}+ \left(v\frac{dv}{dt}-\frac{dv}{dt}v \right)\vec{n}\right]\frac{1}{v^2}=\frac{1}{v}\frac{(d\vec{v})_\perp}{dt}.
\label{dn}
\end{eqnarray}
Note that $d\vec{n}/dt$ is perpendicular  to $\vec{n}$, as expected.

Collecting  the results (\ref{Fr1}) and (\ref{Fp1}) one can write them in the concise form:
\begin{equation}
\vec{F}_{R}\equiv (\vec{F}_{R}){_\parallel}+(\vec{F}_{R}){_\perp}=\frac{dp}{dt}\vec{n}+\gamma{p}\frac{d\vec{n}}{dt},
\label{suma}
\end{equation}
where $\gamma=1/(1-v^2/c^2)^{1/2}$.

If we introduce a vector of relativistic momentum:
\begin{equation}
\vec{p}=p\vec{n}=\frac{m\vec{v}}{\sqrt{1-v^2/c^2}},
\label{ped}
\end{equation}
 which the time derivative can be written as:
\begin{equation}
\frac{d\vec{p}}{dt}=\frac{d(p\vec{n})}{dt}  =\frac{dp}{dt}\vec{n}+p\frac{d\vec{n}}{dt}\equiv \left(\frac{d\vec{p}}{dt}\right)_\parallel+\left(\frac{d\vec{p}}{dt}\right)_\perp,
\label{dp}
\end{equation}
the result (\ref{suma}) of the action of the force $\vec{F}_R$ may be expressed as:
\begin{equation}
\vec{F}_R=\left(\frac{d\vec{p}}{dt}\right)_\parallel+\gamma\left(\frac{d\vec{p}}{dt}\right)_\perp.
\label{zobrotem}
\end{equation}
The last equation may be rewritten in the commonly used in the standard relativity form of the \emph{relativistic equation of motion}:
\begin{equation}
\vec{F}=\frac{d\vec{p}}{dt},
\label{last}
\end{equation}
where $\vec{F}$ is called the force in the frame $S$ and is defined as follows:
\begin{equation}
\vec{F}\equiv \left((\vec{F}_R)_\parallel,\frac{(\vec{F}_R)_\perp}{\gamma}\right).
\label{last'}
\end{equation}
Eq. (\ref{last'}) is the correct (known from the standard approaches) relation between the force $\vec{F}_R$ in the rest frame $R$ and the force  $\vec{F}$ measured in the system of reference $S$.

In this way we have arrived at the desired relativistic
 equation of motion (\ref{last}). Our method shows that if one considers an accelerating body moving in Minkowskian space-time geometry, the form of the equation of motion indispensably must be given by Eq. (\ref{last}). Moreover,  the relativistic momentum defined in Eq. (\ref{ped}) appears to be an inherent compound of this equation. 

Since the frame $S$ is chosen completely arbitrarily, it straightforwardly follows also that the relativistic 
equation of motion (\ref{last}) is the same for \emph{any} frame of reference.  Its \emph{form} remains invariant 
for any inertial system of coordinates. However, in each frame the force is different and 
depends on the velocity $\vec{v}$ the body has just in this frame. Eq. (\ref{last'}) shows that not only the 
value of $\vec{v}$ is important. The \emph{direction} of $\vec{v}$ determines the decomposition of
 force $\vec{F}_R$ on the directions parallel and perpendicular to $\vec{v}$ by means of which the force $\vec{F}$ in the frame $S$ is defined.

\section{A new method of force transformation}

Eq. (\ref{last'}) offers us transformation of force $\vec{F}_R$ from the instantaneous rest frame co-moving with a body to some
 inertial reference frame $S$.
From the practical point of view it is advantageous to express the force $\vec{F}$ registered by an observer in the frame $S$ by 
a force $\vec{F}'$ from some other ''{stationary}'' system of reference $S\,'$. The reason is that it is troublesome to establish
 force $\vec{F}_R$ in the rest frame of moving body. Furthermore, some fundamental  formulae describing forces are
 determined in such a system of coordinates, say $S\,'$, in which the \emph{source} of force is at rest. The example is 
 Coulomb's law for the electrostatic interactions that describes the influence of a source-charge \emph{being at rest} on other (possibly moving) charges.

 To cope with this task we may use  Eq. (\ref{last'}) two times: once for a system $S\,'$ and again for the  system $S$. Important is
 that in the both cases Eq. (\ref{last'}) contains the same force $\vec{F}_R$ established in the rest frame $R$. 
Let in a system of coordinates $S\,'$ an accelerated body has a velocity $\vec{w}$ and the  force in this frame is a function
 $\vec{F'}$. According to our reasoning presented in the previous section the equation of motion in the frame $S\,'$ is: 
\begin{equation}
\vec{F'}=\frac{d\vec{p\,'}}{dt'}
\label{f'p'}
\end{equation}
with $\vec{p\,'}=m\vec{w}/(1-w^2/c^2)^{1/2}$; and using Eq. (\ref{last'}) the force $\vec{F}_R$ registered in the rest frame of accelerated body is given by equality:
\begin{equation}
\vec{F}_R= \left((\vec{F'})_{_{\parallel\vec{w}}},\; \gamma_{{w}}{(\vec{F'})_{_{\perp\vec{w}}}}\right),
\label{fnf'}
\end{equation}
Note that the direction parallel and perpendicular are established with respect to the direction of vector $\vec{w}$.

Now, in some other system of coordinates $S$ the same body possesses a velocity $\vec{v}$. Eq. (\ref{last}) is valid and the force measured in this frame is $\vec{F}$. To express $\vec{F}$ by means of $\vec{F'}$  again we refer to Eq. (\ref{last'}). Let us rewrite  Eq. (\ref{last'}) specifying explicitly that the respective directions are related to the velocity $\vec{v}$:

\begin{equation}
\vec{F}= \left((\vec{F}_R)_{_{\parallel\vec{v}}},\;\frac{(\vec{F}_R)_{_{\perp\vec{v}}}}{\gamma_v}\right),
\label{ffn}
\end{equation}
On the basis of Eq. (\ref{fnf'})  $\vec{F}_R$ it is a function of $\vec{F'}$. Thus Eq. (\ref{ffn}) together with Eq. (\ref{fnf'})  represents the desired (implicit) relation between forces $\vec{F}$ and $\vec{F'}$. 

Concluding, the procedure of finding the general relation between forces in different laboratory frames relies on 
the geometrical projections of the
 force $\vec{F'}$ on the directions of the respective velocities. However, although Eqs. (\ref{fnf'}) and (\ref{ffn}) look very simple, using them in particular cases requires some caution and practice.

\section{Example}

 To show how the described procedure of transformation of force gives an explicit equation joining $\vec{F}$ and $\vec{F'}$ let us consider simple but instructive example. It may be used to show that magnetic field can be derived straightforwardly solely from the Coulomb law.
We do not need to introduce neither the electric field transformation  nor  the relativistic effect of contraction leading to the change of density of charge.  The presence of magnetic field  is shown to be caused simply by movement of a charged point particle. Additionally, this example teaches about the Thomas rotation which must be kept in mind while  the forces are transformed.

Let  a charge $Q$ (source of force) be at rest in the frame $S'$. Some other body having a charge $q$ moves in $S'$ directly away in the direction of the $y'$-axis with a velocity $w$ (see Fig. 1). 
\begin{figure}[h!t]
\center{\includegraphics*[width=14cm]{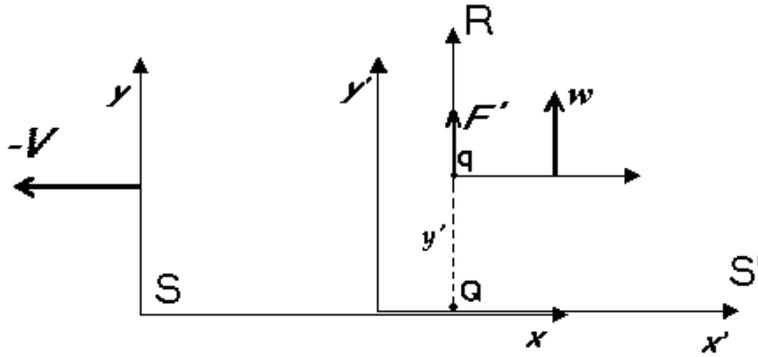}}
\caption{Situation from the point of view of the frame $S'$.}
 \label{Fg1}
\end{figure}

The force $\vec{F'}$ acting on the charge $q$ has a nonzero component only along the velocity $w$,  that is:
\begin{equation}
\vec{F'}=(0,F',0)
\label{f'}
\end{equation}
where $F'=kqQ/y'^2$.

Let the axes of the rest frame $R$ be oriented along the respective axes of the system $S'$. Certainly, also in $R$ the force has only the component along the velocity $w$. From Eq. (\ref{fnf'}) we get the force $\vec{F}_R$ in the rest frame of moving body:
\begin{equation}
\vec{F}_R= (0, F',0).
\label{fn}
\end{equation}

From the point of view of some other system $S$ the frame $S'$ moves along the $x$-axis parallel to the $x'$-axis with a velocity $V$. 
The charge $q$ has then a velocity $\vec{v}$ with respect to the frame $S$. According to Eq. (\ref{ffn}) to get the force $\vec{F}$ we have
 to project the force $\vec{F}_R$ on the directions parallel and perpendicular to $\vec{v}$. We emphasize that this operation must be made 
in the frame $R$, so that we need to know the coordinates of the velocity $\vec{v}$ as measured in this system of reference. First then from
 the well known general equations joining velocities in different systems  we can find the velocity $-\vec{v}$ of the frame $S$ as registered in 
$R$. Since the relative velocity of frame $S'$ with respect to the frame $R$ is $-w$ and the frame $S$ has in the system $S'$ the velocity $-V$
 perpendicular to $-v$, we have [3]:
\begin{equation}
(-v)_{_{\parallel w}}=-w,\;\;\; \;\;\;   ( -v)_{_{\perp w}}=-V/\gamma_w
\label{-w1}
\end{equation}
The relative velocity $\vec{v}$ the system $R$ possesses with respect to the system $S$ has then in the system $R$ the components:
\begin{equation}
\vec{v}=\left(\frac{V}{\gamma_w}, w,0\right)_R.
\label{wR}
\end{equation}
\begin{figure}[h!t]
\center{\includegraphics*[width=14cm]{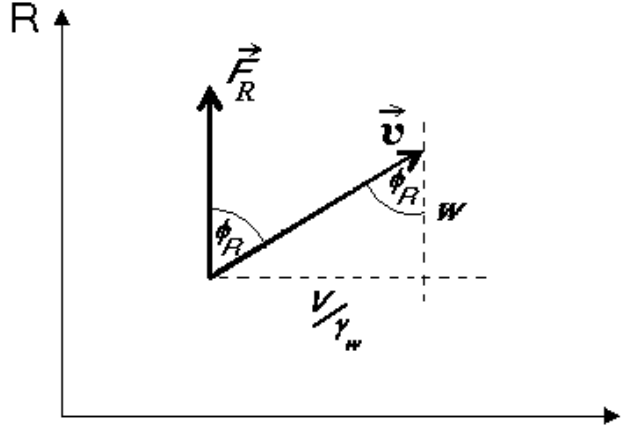}}
\caption{Vector of the relative velocity $\vec{v}$ between the frames $R$ and $S$ as seen in the system $R$.} 
\label{Fg2}
\end{figure}

From Fig. 2 and with help of Eq. (\ref{wR}) we see that:
\begin{equation}
(F_R)_{_\parallel v}=F_R\cos\phi_R=F_R\frac{w}{v}\\,
\label{fninR}
\end{equation}
\begin{equation}
(F_R)_{_\perp v}=F_R\sin\phi_R=F_R\frac{V}{v\gamma_w}.
\label{fninR'}
\end{equation}

Now, from Eqs (\ref{fninR}) and (\ref{fninR'}) and according to Eq. (\ref{ffn}) the respective components of the force $\vec{F}$ measured in $S$ parallel
 and perpendicular to $\vec{v}$ are:
\begin{equation}
F_{_\parallel v}=F_R\frac{w}{v},
\label{finS}
\end{equation}
\begin{equation}
F_{_\perp v}=F_R\frac{V}{v\gamma_w\gamma_v}.
\label{finS'}
\end{equation}
More convenient is to have this force written by means of components along the axes of the system $S$. First then
 we have to find the respective components of the velocity $\vec{v}$ in the frame $S$. Because $S'$ moves with 
respect to $S$ with the velocity $V$  and the velocity of body $w$ measured in $S'$ is perpendicular to $V$, the velocity in the frame $S$ is [3]:
\begin{equation}
v_x\equiv v_{_{\parallel V}}=V,\;\;\;\; \;\;\;    v_y\equiv v_{_{\perp V}}=w/\gamma_V
\label{w}
\end{equation}
or
\begin{equation}
\vec{v}=\left(V,\frac{w}{\gamma_V},0\right)_S.
\end{equation}
Note that the same relative velocity $\vec{v}$ has different components in the frames $R$ and $S$ (compare to Eq. (\ref{wR})). This effect is called
 the Thomas rotation [4, 5] and it means that the systems $R$ and $S$ appear to be rotated one with respect to the other (see Fig. 3). 
\begin{figure}[h!t]
\center{\includegraphics*[width=14cm]{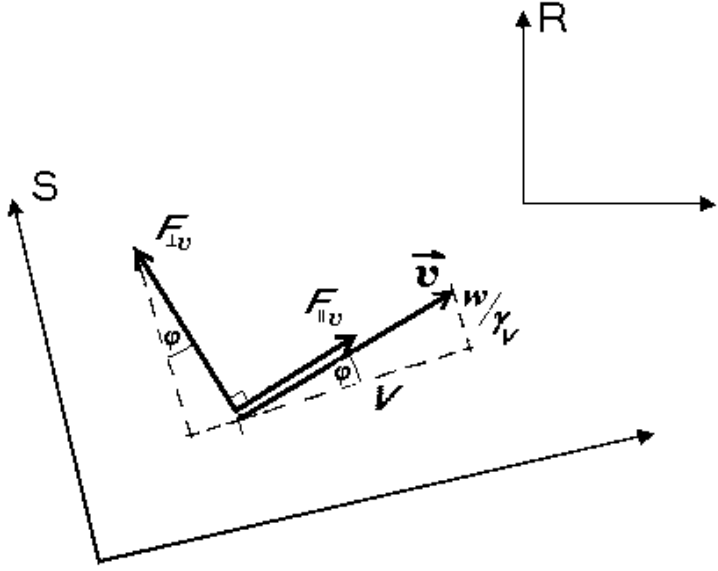}}
\caption{Vector $\vec{v}$ and the respective components of the force $\vec{F}$ in the frame $S$.}
 \label{Fg3}
\end{figure}

According to Fig. 3,  Eqs (\ref{finS}), (\ref{finS'}) and (\ref{w}) we have:
\begin{equation}
F_x= F_{_\parallel v}\cos\phi-F_{_\perp v}\sin\phi=F_R\frac{wV}{v^2}-F_R\frac{Vw}{v^2\gamma_w\gamma_v\gamma_V}=\frac{F_RwV}{v^2}\left(1-\frac{1}{\gamma_v\gamma_w\gamma_V}\right).
\label{fx1}
\end{equation}
From Eq. (\ref{w}) we get:
\begin{equation}
v^2=V^2+\frac{w^2}{\gamma^2_V}
\label{wVv}
\end{equation}
One can then obtain after simple algebra that:
\begin{equation}
\gamma_v=\gamma_w\gamma_V
\label{gammy}
\end{equation}
and also from the definition of $\gamma$ we have:
\begin{equation}
\frac{v^2}{c^2}=\frac{\gamma^2_v-1}{\gamma^2_v}
\label{beta}
\end{equation}
Inserting Eqs (\ref{gammy}) and (\ref{beta}) into Eq. (\ref{fx1}) and remembering that $F_R=F'$ we get:
\begin{equation}
F_x=F'\frac{wV}{c^2}
\label{fx}
\end{equation}
Similarly in an analogous way one can obtain that:
\begin{equation}
F_y=\frac{F'}{\gamma_V}
\label{fy}
\end{equation}
Because $y'=y$, we have $F'=kQq/y^2$. In this way the force $\vec{F}$ may be expressed solely by variables used in $S$:
\begin{equation}
\vec{F}=\left(\frac{kQq}{y^2}\frac{\gamma_V v_yV}{c^2},\frac{kQq}{\gamma_Vy^2},0\right),
\label{sila}
\end{equation}
where we have used the relation (\ref{w}) and substituted $w=\gamma_V v_y$.

The result (\ref{sila}) is surprising. While in the frame $S'$ the force has only one component along the $y'$-axis (Eq. (\ref{f'})), in the system $S$ (having its axes pointed in the same direction as the axes of the frame $S'$) there appears additionally a component along the $x$-axis.
We recognize it as the magnetic part of the Lorentz force:
\begin{equation}
F_x=qBv_y,
\end{equation}
 where the factor $B$ stands here for:
\begin{equation}
 B=kQV\gamma_V/y^2c^2. 
\end{equation}
It is the correct 
formulae on the magnetic induction satisfying the relativistically invariant Maxwell equations. For a small velocity $V$ (i.e. with $\gamma_V\approx1$) it becomes the Biot-Savart expression on the magnetic induction produced in the frame $S$ at the location of the charge $q$ by the moving with velocity $V$ charge $Q$. 

In turn, the $y$-component of the force $\vec{F}$ given by Eq. (\ref{sila}) has a form $F_y=qE_y$, where $E_y={kQq}/{\gamma_Vy^2}$ represents
 the electric field in the frame $S$. Note that it is properly transformed field $E'_y={kQq}/{y'^2}$ from the frame $S'$ to the frame $S$, i.e. 
$E_y=E'_y/\gamma_V$. Finally then, if we ascribe a vector character to the field $B$ and assume it is a vector pointing along the $z$-axis of 
the frame $S$, Eq. (\ref{sila}) may be rewritten as:
\begin{equation}
\vec{F}=q\vec{E}+q\vec{v}\times\vec{B},
\end{equation}
which shows that Coulomb's force $\vec{F'}$ after the Lorentz  transformation has in the frame $S$ the form of the Lorentz force. 
This result is completely in accordance with the outcomes obtained by other authors by means of different general methods [6-8].

\section*{References}

\noindent $[1]$ G. N. Lewis, R. C. Tolman, "The principle of relativity and non-Newtonian mechanics," Phil. Mag. \textbf{18}, 510-523 (1909).\\
$[2]$ R. C. Tolman, "Non-Newtonian mechanics, the mass of a moving body," Phil. Mag. \textbf{23}, 375-380 (1912).\\
$[3]$ J. D. Jackson, \emph{Classical Electrodynamics} (Willey, New York, 1999), 3rd ed., Sec. 11.4.\\
$[4]$ J. P. Costella, B. H. McKellar, A. A. Rawlinson, G. J. Stephenson, Jr., "The Thomas rotation," Am. J. Phys. \textbf {69}, 837-847 (2001).\\
$[5]$ K. R\c {e}bilas, "Comment on "The Thomas rotation," by J. P. Costella \emph{et al.}" Am. J. Phys. \textbf{70}, 1163-1165 (2002).\\
$[6]$ W. G. V. Rosser, \emph{Classical Electromagnetism via Relativity} (Butterworths, London 1968), Sec 3.\\
$[7]$ D. H. Frish, L. Wilets, "Development of the Maxwell-Lorentz Equations from Special Relativity and Gauss's Law", Am. J. Phys. \textbf{24}, 574-579 (1956).\\
$[8]$ K. R\c {e}bilas, ''A way to discover Maxwell's equations theoretically'', Found. Phys. Lett. 19 (4), 337-351 (2006).\\ 
\end{document}